\documentclass[twocolumn,times,tighten]{aastex63}

\newcommand{\pion}[2]{{#1}\,{\sc {#2}}}
\newcommand{\fion}[2]{[{#1}\,{\sc {#2}}]}
\newcommand{\vunit}{\mbox{\,\,km\,s$^{-1}$}}
\newcommand{\mic}{\mbox{$\,\,\mu$m}} 
\newcommand{\Msun}{\mbox{\,\,M$_\odot$}}
\newcommand{\Lsun}{\mbox{\,\,L$_\odot$}}
\newcommand{\nucl}[2]{\mbox{$^{#1}${#2}}}

\usepackage{color}
\usepackage{natbib}

\usepackage{apjfonts}
\usepackage{graphicx}

\shorttitle{Infrared spectroscopy of CK Vulpeculae}
\shortauthors{D. P. K. Banerjee et al.}
\begin{document}
\title{NEAR-INFRARED SPECTROSCOPY OF CK VULPECULAE: \\
REVEALING A REMARKABLY POWERFUL BLAST FROM THE PAST}
\author{D. P. K. Banerjee}
\affiliation{Astronomy \& Astrophysics Division, Physical Research Laboratory, Navrangpura, Ahmedabad, 380009, India}
\email{dpkb12345@gmail.com}
\author{T. R. Geballe}
\affiliation{Gemini Observatory/NSF's NOIRLab, 670 N. A'ohoku Place, Hilo, Hawai'i, 96720, USA}
\author{A. Evans}
\affiliation{Astrophysics Group, Lennard Jones Laboratories, Keele University,
Keele, Staffordshire, ST5 5BG, UK}
\author{M. Shahbandeh}
\affiliation{Department of Physics, Florida State University, 77 Chieftain Way, Tallahassee, FL 32306-4350, USA}
\author{C. E. Woodward}
\author{R. D. Gehrz}
\affiliation{Minnesota Institute for Astrophysics, School of Physics \& Astronomy, 116 Church Street SE, University of Minnesota, Minneapolis, MN 55455, USA}
\author{S. P. S. Eyres}
\affiliation{Faculty of Computing, Engineering, and Science, University of 
South Wales, Pontypridd, CF37 1DL, UK}
\author{S. Starrfield}
\affiliation{School of Earth and Space Exploration, Arizona State University, Box 871404, Tempe, AZ 85287-1404, USA}
\author{A. Zijlstra}
\affiliation{Jodrell Bank Centre for Astrophysics, School of Physics and Astronomy, University of Manchester, Manchester M13 9PL, UK}

\begin{abstract}

CK Vulpeculae, which erupted in AD 1670-71, was long considered
to be a nova outburst; however, recent observations have required that
alternative scenarios be considered. Long slit infrared spectroscopy of a 
forbidden line of iron reported here has revealed high line-of-sight velocities
($\sim\pm900$\vunit) of the ansae at the tips of the bipolar lobes 
imaged in H$\alpha$ in 2010. The deprojected  velocities of the tips 
are approximately $\pm2130$\vunit\ assuming the previously derived inclination angle 
of $65^\circ$ for the axis of cylindrical symmetry of the bipolar nebula. Such high 
velocities are in stark contrast to previous reports of much lower expansion velocities 
in CK~Vul. Based on the deprojected velocities of the tips and their angular 
expansion measured over a 10-year baseline, we derive a revised estimate, 
with estimated uncertainties, of $3.2^{+0.9}_{-0.6}$~kpc 
for the distance to CK~Vul. 
This implies that the absolute visual magnitude at the peak of the 1670 
explosion was $M_V = -12.4^{+1.3}_{-2.4}$,
indicating that the 1670 event was far more luminous than previous estimates 
and brighter than any classical nova or any Galactic stellar merger.
We propose that CK~Vul belongs to the class of 
Intermediate Luminosity Optical Transients (ILOTs), objects which bridge 
the luminosity gap between novae and supernovae.  While eruptions in 
lower luminosity ILOTs are attributed to merger events, the origin of 
the highly luminous ILOT outbursts is 
currently not known.
\end{abstract}

\keywords{stars: individual (CK~Vul)
--- stars: peculiar --- stars: winds, outflows
---infrared: stars}

\section{Introduction}
\label{Intro}
CK Vulpeculae (long referred to as Nova Vulpeculae 1670) was believed 
until recently to possibly be the earliest documented nova. It was 
discovered as a naked-eye object on 1670 June 20 by Anthelme, and
independently on July 25 by Hevelius. Reconstruction of its light 
curve \citep{shara85} shows it had a visual brightness maximum of 
approximately 3~mag at discovery, followed successively by 
a fading and a second maximum at approximately 2.6~mag in 
1671 March. It subsequently faded from naked-eye view, although a 
final maximum of $5.5-6$~mag was observed by 
Hevelius in 1672 March. Since 1679, attempts to recover the  object 
were unsuccessful until the discovery in 1982 of a small patch of H$\alpha$ 
nebulosity at the site of the 1670 event \citep{shara82}.
A thorough discussion is given by \cite{shara85}.

\begin{figure}[]
\includegraphics[width=8cm]{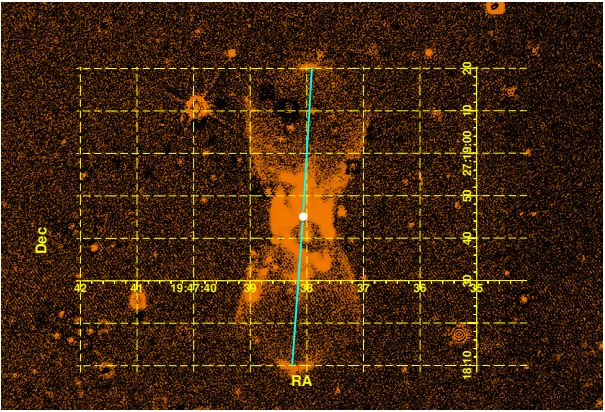}
\caption{H$\alpha$ continuum-subtracted image of CK~Vul observed on 2010 
June 22 with the Gemini North Telescope (image kindly provided by M. Hajduk) . 
The position of the compact radio source is indicated by the white dot. 
The slit (cyan line) was centered on this source and passes through the tips of the 
ansae located near the top and bottom center of the image.}
\label{neb}
\end{figure}

Submillimeter photometry of CK~Vul \citep{evans02} revealed a large 
excess due to emission from dust, inconsistent with CK~Vul being a 
classical nova remnant.  A faint $70''$ bipolar nebula, oriented almost exactly N-S 
(Figure~\ref{neb}), was discovered in H$\alpha$ imaging by \cite{hajduk07}, 
who also reported a compact radio source located at the point of symmetry. 
Infrared (IR) spectroscopy \citep{evans16} revealed a rich 
spectrum of hydrocarbons, together with H$_2$, HCN and metallic fine structure line 
emission. A $\sim10''$ dusty bipolar submillimeter structure, symmetric
about a small disk-like feature, the center of which coincides with the compact 
radio source, was reported by \cite{eyres18}. This structure,
which aligns with dark lanes in the H$\alpha$ image, also shows
emission from several small organic molecules. 

The central star of CK~Vul is deeply enshrouded in dust, and
is visible in neither the optical nor the near-IR.  The current identification 
of CK Vul in the SIMBAD database {\em is incorrect}.
In this study, we adopt the radio coordinates \citep{hajduk07} listed in Table 1
for the central source of CK Vul.

No consensus exists on the cause of the 1670 eruption. 
Proposed scenarios include a nova eruption, various types of stellar merger,
and a very late thermal pulse in a post-AGB star
\citep{shara82,shara85,hajduk07,kaminski15,kaminski17,kaminski18a,eyres18}.
A kinematic study of the nebula could be a means of discriminating 
between some of the above mechansims. For example, stellar mergers 
rarely produce outflows with velocities exceeding
500\vunit. On the other hand, nova eruptions can generate ejecta velocities 
of up to several thousand \vunit. 

Here we present near-IR spectra of CK~Vul, obtained through a slit 
positioned as shown in Figure \ref{neb}. The orientation 
was chosen so that the slit intercepted the radio source at the center of the bipolar 
structure and the northern and southern
tips of the ansae, approximately $36''$ from the center, where the expansion 
velocities are expected to be largest in a homologous or Hubble flow 
\citep[][and references therein]{balick2002}. \\


\section{Observations and Data Reduction}

Spectroscopy of CK~Vul was obtained at the Frederick C. Gillett Gemini 
North Telescope on Maunakea on the nights of 2020 31 August and 
1 September (UT), using the Gemini Near-InfraRed Spectrograph 
\citep[GNIRS;][]{elias06}. Sky conditions were excellent throughout 
the observations. The spectrograph configuration included its $99''$ 
slit oriented $3^\circ$ west of north, a slit width of $0.45''$, and 
the short focal length blue camera, which provides a plate scale of 
$0.15''$/array pixel. The $110.5\,\,\ell$/mm grating 
in GNIRS was set to a central wavelength of 1.26\mic, which gave a 
coverage of 1.203$-$1.317\mic\ and a resolving power, 
$R$, of 4800 (corresponding to resolutions of 0.00026\mic\ and 62\vunit).

Individual exposure times were 300 seconds. The total exposure times 
on source were 1800 seconds on 31 August and 2400 seconds on 1 September. 
The same exposure times were used at a location $46''$ to the east, 
selected to minimize the number of stars in the ``sky'' slit. Exposures 
were obtained in a the standard ABBA manner. The star HIP 98699 (A1V) 
was observed immediately following CK~Vul; the airmass match was within 
0.04 airmasses on each night.

On both nights the slit was centered on the radio coordinates (Table~1) using a blind offset 
from a $K\approx10$ mag star located $0.6'$ distant. The offsetting accuracy of Gemini 
is considerably less than a GNIRS pixel. Although CK~Vul itself 
was not detected in the near-IR, we are confident of the angular 
distances from it to the spectral features reported in this paper.

Examination of raw sky-subtracted frames revealed within $10''$ of 
CK~Vul several patches of emission along the slit in the forbidden 
line of \pion{Fe}{ii} at 1.25702\mic\ (vacuum). The \pion{H}{i}  Pa$\beta$ line 
at 1.28216\mic\ was also weakly present in a portion 
of this region within a few arcseconds of the radio source. 

Prompted by the report by \cite{hajduk13} of H$\alpha$ line emission 
with high proper motions at the northern and southern tips of the bipolar 
nebula, we searched the spectral images on both nights and found 
regions of faint and highly Doppler-shifted ($\sim\pm900$ km s$^{-1}$)
emission in the \fion{Fe}{ii} line, about one arcsecond in extent, close to the
locations of the H$\alpha$ tips.
The northern emission stands out fairly well. The southern 
emission has superimposed on part of it the spectrum of a faint star.
However, careful data reduction clearly established the emission's existence.
We have considered  whether these components are lines of some other atomic species,
but think it highly unlikely and unreasonable that an atomic species would selectively 
create only one-sided emission (i.e. emission in only one of the two ansae) and also
that emission would appear only in the tips and not in the bright central region 
where it would be expected to be strong. Thus, we are confident in our identification.

Full reduction of the spectra from each night was limited to the 
wavelength interval containing the \fion{Fe}{ii} and Pa~$\beta$ 
lines, there being no detected lines outside of that
interval. It involved the standard steps of flat-fielding, spatial 
and spectral rectification of the images, removal of the effects 
of cosmic ray hits, wavelength calibration using an argon lamp, 
and binning of the spectra, in this case into intervals of 
0.0002\mic\ (48\vunit).

Spectra of the twelve regions where the \fion{Fe}{ii} line was detected 
were extracted from the rectified CK~Vul reduced image from each night. 
Each was flux-calibrated by dividing by the spectrum of the standard star, 
including correcting for slit loss in the spectrum of the star. The 
spectra from the two nights were then combined to produce a final 
spectrum for each region. Pertinent information for each spectrum
is provided in Table~\ref{FeII}.

\begin{table*}
\begin{center}
\caption{Observed [Fe\,{\sc ii}] 1.257\mic\ line.\label{FeII}}
\begin{tabular}{lcccc}\hline\hline
Offset$^a$   & Aperture         &  Vel. of peak$^b$     &	Flux$^c$ & Notes \\
arcsec &  arcsec  &         km s$^{-1}$ hel.          &    10$^{-19}$ W m$^{-2}$        &         \\\hline

+36.7 &  0.45 $\times$ 0.90  &  +869  & 1.39  & blue shoulder \\
+7.9  &  0.45 $\times$ 0.75  &  -111  & 0.53  & \\
+4.2  &  0.45 $\times$ 0.75  &  -147  & 1.48  & blue shoulder \\
+4.2  &  0.45 $\times$ 0.75  &  -40   & 2.15  & \\
+2.9  &  0.45 $\times$ 0.75  &  -118  & 6.25 & \\
+1.5  &  0.45 $\times$ 0.75  &  -106  & 5.53  &  Pa $\beta$ at $-$120 km s$^{-1}$\\
+0.6  &  0.45 $\times$ 0.45  &  -300  & 0.84  & \\
+0.6  &  0.45 $\times$ 0.45  &  -63   & 2.62  & Pa $\beta$ at $-$50 km s$^{-1}$ \\
0.0     &  0.45 $\times$ 0.45  &  -302  & 1.16  &  \\
0.0     &  0.45 $\times$ 0.45  &  -47   & 0.56  & Pa $\beta$ at $-$50 km s$^{-1}$ \\
-0.6  &  0.45 $\times$ 0.45  &  -285  & 0.25  & \\
-3.3  &  0.45 $\times$ 0.75  &  +68   & 2.25  & \\
-4.1  &  0.45 $\times$ 0.75  &  +20   & 3.50  & \\
-8.9  &  0.45 $\times$ 0.75  &  +137  & 2.05  & \\
-37.1 &  0.45 $\times$ 1.20  &  -922  & 0.84  & may have red shoulder\\
\hline \hline
\end{tabular}
\end{center}
$^a$ From radio coordinates of CK Vul ($J2000$ RA =19:47:38.074, Dec = +27:18:45.16) along slit
at position angle $3\deg$ west of north. \\
$^b$ Uncertainty $\sim$10 km s$^{-1}$. \\
$^c$ Typical random uncertainty 0.10 W m$^{-2}$. Uncertainty in absolute flux calibration $\sim$20 percent.
\end{table*}

\begin{figure}[]
\includegraphics[width=0.45\textwidth]{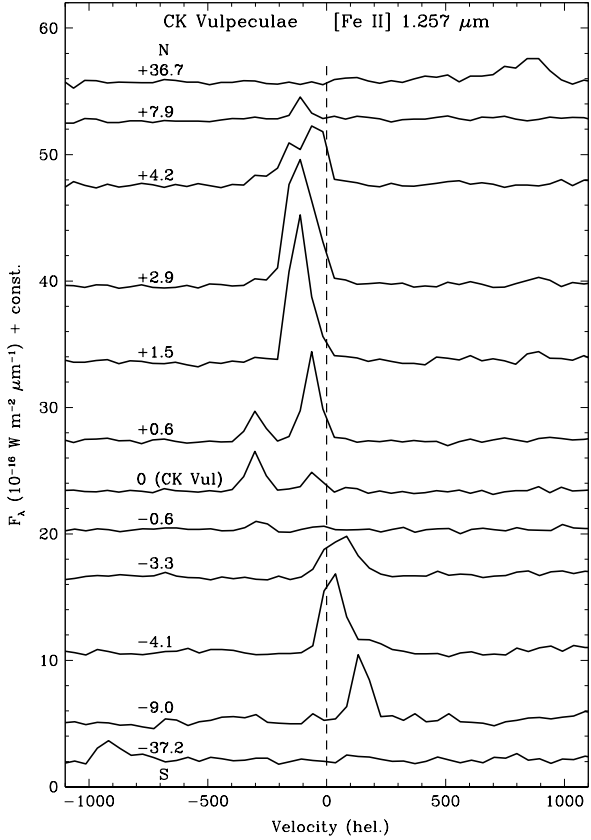}
\caption{Spectra of the \fion{Fe}{ii} 1.257\mic\ line in the CK Vul nebula
at the twelve locations along the slit where the line was detected. 
See Table 1 for details.}
\label{swave}
\end{figure}

\begin{figure}[]
\includegraphics[width=0.45\textwidth, angle=0]{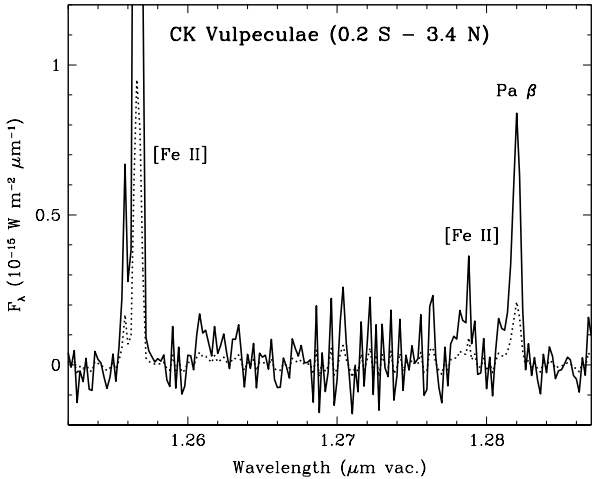}
\caption{Spectrum extracted over a 3\farcs6 region near the center, 
showing three lines (\fion{Fe}{ii} 1.257\mic\ and 1.279\mic,
and  \pion{H}{i} Pa$\beta$ 1.282\mic). Dotted line: same spectrum scaled 
down by a factor of four.}
\label{fe-all}
\end{figure}

\newpage
\section{Results}

Spectra of \fion{Fe}{ii} 1.257\mic\ at all positions 
where the line was detected are presented in Figure~\ref{swave}. 
Emission was also detected in the Pa$\beta$ line, but only
 close to the radio source. A third line, 
\fion{Fe}{ii} at 1.279\mic\,, was also weakly detected, but only in 
a few arcsecond long segment of the slit where the 1.257~$\mu$m line is at its
brightest. Figure~\ref{fe-all} shows the spectrum of that portion of the central 
region where all three lines are present. 

As can be seen in Figure~\ref{swave}, the  \fion{Fe}{ii} 1.257\mic\ line
emission is prominent at various locations within $\pm9''$ of the central
source. The velocity pattern within this inner region is complex, with large 
variations with position and at some locations more than one component 
present. Discussion and modeling of the velocity field will be 
reported elsewhere.
 
The highlight of these observations is the high line-of-sight velocities
observed in the tips of the north and south ansae in the \fion{Fe}{ii} line, 
peaking at $\sim+870$\vunit\  and $\sim-920$\vunit, respectively.  
\cite{hajduk13}, using the morpho-kinematic modelling tool {\sc shape} 
\citep{steffen06} found that that the inclination angle of the bipolar H$\alpha$ 
nebula is  $65^\circ$ relative to the line-of-sight. That implies that
the northern and southern tips are moving outward from the center at 
2130\vunit~~(900\vunit/cos $65^\circ$). This is vastly higher than previously reported 
expansion velocities in CK Vul, which are of order a few hundred km s$^{-1}$ 
or less \citep{shara85, hajduk07}.
 
The reason for these previously reported low velocities is now clear: all of them 
are for regions located around the central waist of the hourglass 
nebula (i.e. the equatorial region) where low expansion velocities 
are expected. Hourglass or bipolar nebulae are believed to be 
generated when the outflowing ejecta from the central source are 
constricted by a material over-densities in the equatorial 
planes, i.e. at the waist of the hourglasses. The ejecta expand freely and rapidly in 
the polar directions, but slowly in the equatorial region. 
High velocities are expected in the polar regions by virtue of the 
larger distance to the tips of the ansae.

For the H$\alpha$ frame of 2010 June 22 (Figure~\ref{neb}) we measure 
the angular separation between the centers of the tips of the ansae at the positions 
where the GNIRS slit intersected them to be 71\farcs29$\pm$0\farcs20. 
Using an accurate value of 0\farcs1517$\pm$0\farcs0002~pixel$^{-1}$ for 
the plate scale of GNIRS in the configuration in which it was used, 
we find an angular separation of 73\farcs87$\pm$0\farcs23 for the  
centers of the tips in the \fion{Fe}{ii} 1.257\mic\ emission line. Because 
\fion{Fe}{ii} emission arises in a region where hydrogen is partially ionized
\citep{mouri00}, we assume that H$\alpha$ and \fion{Fe}{ii} 
are co-spatial. An expansion of 2\farcs58$\pm$0\farcs30 has therefore 
occurred over a 10.2~year period (2010 June 22 to 2020 August 31). 
Assuming constant velocity this implies that the expansion 
began $290\pm40$ years ago. Comparison with the the likely date of the 
ejection event, 1670 or 1671, suggests that the deceleration of the 
tips, if any, has been rather modest.

Assuming a constant transverse velocity of 1930\vunit\ (2130 sin $65^\circ$) 
between 2010 and 2020  (i.e., no significant deceleration of the ejecta), 
the distance to CK~Vul is 3.2~kpc, independent 
of reddening. {\em This  distance is a factor of 4.5-6 times larger than 
other estimates} \citep{shara85,hajduk13}. 

The uncertainty in the derived inclination angle was not given 
by \cite{hajduk13} and is not incorporated in the above distance 
uncertainty. At high inclination angles, expansion velocity, distance, 
and absolute magnitude all have non-linear dependences on this angle.
An uncertainty of $5^\circ$ in the assumed $65^\circ$ 
gives the distance to CK Vul as $3.2^{+0.9}_{-0.6}$~kpc; further allowing for the
uncertainty in reddening gives the absolute visual magnitude at outburst
as  $M_V=-12.4^{+1.3}_{-2.4}$.

The above distance of $3.2^{+0.9}_{-0.6}$~kpc is in apparent conflict with distance
estimates by \cite{hajduk13} of $\sim$500 pc and $\sim$2 kpc to two variable field
stars, which they argued lie behind some of the ejecta of CK Vul, which has 
caused their recent brightness fluctuations and apparent Li overabundances. 
However, the former of these has a $Gaia$ parallax \citep{bai18} of 4.9 kpc  
and thus no longer conflicts. The latter star is too faint for $Gaia$ and a more 
precise distance would need to be determined by some other means.

\section{Discussion}

\subsection{Line emission in the central few arcseconds}

Here we confine our discussion to the central region and the spectrum in  
Figure~\ref{fe-all}, in which all three lines (\fion{Fe}{ii} 1.257 and 1.279\,\mic, and 
Pa$\beta$) are present. This spectrum includes contributions from
several disparate areas. Thus the properties derived here are mean values 
and do not necesaarly represent any individual region. Based on the 
strength of \fion{Fe}{ii} 1.257\mic\ relative to Pa$\beta$, we assume
 that the gas in this region is largely collisionally excited. However, 
the presence of a compact radio source \citep{hajduk07} indicates the 
possibility of some photoionization.
 
As discussed by \cite{mouri00}, the \fion{Fe}{ii} emission
arises in a region in which hydrogen is partially ionized.
The critical density for collisional de-excitation of the upper level of the 
1.257\mic\ line is \[n_{\rm crit}  = 5.6\times10^4 \:\: 
\left ( \frac{T_e}{10,000~\mbox{K}} \right)^{0.66}\:\:\mbox{~cm$^{-3}$} ,\]
where $T_e$ is the electron temperature. Useful diagnostics are the
 \fion{Fe}{ii}~1.257\mic\,/\,Pa$\beta$ and 
\fion{Fe}{ii}~1.279\mic\,/\,\fion{Fe}{ii}~1.257\mic\
flux ratios, which are $4.9\pm0.6$ and $0.07\pm0.02$ respectively.

The \fion{Fe}{ii} 1.257\mic\ / \,Pa$\beta$ flux ratio suggests shock 
velocities of $\sim70\pm5$\,\vunit\, marginally below the 
threshold (75\vunit) for an ionizing shock \citep{mouri00}. 
Note that this value is for solar abundances, which may not be 
applicable to these ejecta. The  
\fion{Fe}{ii}~1.279\mic\,/\,\fion{Fe}{ii}~1.257\mic\ flux ratio
indicates $T_e\simeq6000$~K, although 
the deduced $T_e$ is virtually independent of the flux ratio.
This temperature gives a critical density of $n_{\rm crit} 
\simeq 4\times10^4$~cm$^{-3}$ for the upper level of the 
1.257\mic\ line.

\subsection{The outer tips}

The morphologies of the \fion{Fe}{ii} 1.257\mic\ tips in the 
CK~Vul nebula closely resemble the ``fingertips" observed in the 
Orion Molecular Cloud by \citet[][see e.g., their Fig.~9]{bally15} in 
the shock-excited forbidden \fion{Fe}{ii} 1.644\mic\ line. 
There over one-hundred high density clumps of gas are being ejected 
from the cloud at typical speeds of 300\vunit, which \citeauthor{bally15}
concluded is the result of a violent event $\sim$500 yr ago 
during the formation of a dense cluster of massive stars.

We propose that excitation of the upper level of the \fion{Fe}{ii} 
1.257\mic\ line in the CK~Vul tips is the result of the same 
phenomenon as that proposed by \cite{bally15} for the  \fion{Fe}{ii} line
in Orion. In each of the tips a low velocity reverse shock 
is being driven into  a high density gas clump ejected from CK~Vul 
as each clump encounters much lower density interstellar gas. 
The shock ionizes and collisionally excites the Fe and, at least to 
some extent, the hydrogen. It is possible that lines of molecular 
hydrogen are also associated with these clunps, as has been observed 
in detail in Orion by \cite{tedds1999}. Observations of them could 
provide constraints on the physical properties of the clumps.

\subsection{The nature of CK~Vul}

The greatly increased distance to CK Vul derived here has dramatic
implications for the dust mass and luminosity in the bipolar
nebula and for the luminosities of the explosive events in 1670-71. 
First, the dust mass in the compact bipolar dusty nebula reported by \cite{eyres18} now
becomes $4.3\times10^{-3}$\Msun, of which $3.3\times10^{-3}$\Msun\
lies in the diffuse emission, and $1.0\times10^{-3}$\Msun\ lies in
the disk. The {\em present} luminosity of the central
object, as determined from the spectral energy distribution of the 
dust is 20\Lsun\ \citep[see][]{kaminski15}. 

Second, assuming the maximum visual brightness $m_{V}$ in 1671 was 
$2.6\pm0.3$ \citep{shara85}, and that the visual extinction $A_{V}$ 
in the direction of CK~Vul at a distance $d = 3.2$~kpc is 
$2.47\pm0.45$ \citep{marshall06}, the absolute magnitude of CK Vul 
at maximum was $-12.4^{+1.3}_{-2.4}$. 
It is thus clear that {\em CK~Vul 
underwent a much more luminous sequence of explosions 
in 1670-71 than hitherto recognized.} At peak its absolute magnitude 
was at least 3~mag brighter than the brightest Galactic classical nova, which
generally have $M_{V}$ at maximum in the range --6.5 to --9.5 
\citep[see e.g.][]{CN2}. A few luminous extragalactic novae 
have been known to approach  $M_{V} = -10$ \citep[e.g.][]{kasliwal11b}.

CK~Vul might be considered as an extreme example of a ``Luminous 
Red Nova'' \citep[LRN;][]{kulkarni07}. Thus, it is of interest to 
compare its peak nebular expansion velocity of $\sim2000$\vunit\ 
with those of others of that class, such as V1309~Sco, V4332~Sgr 
and V838~Mon. V4332~Sgr produced an 
H$\alpha$ line with full width at zero intensity $\sim550$\vunit\ \citep{martini99}, 
a far lower expansion velocity than CK~Vul. Similarly, 
the P Cygni profiles in an early spectrum of V838 Mon 
\citep{lynch04} have separations between their absorption and emission peaks of 
$540\pm140$\vunit. Recent millimeter/submillimeter-wave observations
of V4332~Sgr, V838~Mon, and V1309~Sco  taken 22, 14,
and 8~yr, respectively, after their eruptions \citep{kaminski18b},
reveal molecular emission lines having 
full widths not exceeding 400\vunit. In the few published spectra of 
extragalactic LRNe (e.g  M31~RV and PTF10fqs) also only narrow 
emission lines are present, indicating velocities similar to 
those of their Galactic counterparts \citep{rich89,kasliwal12}. 
In the case of PTF10fqs, in addition to the  
narrow Balmer and calcium emission features, there is a 
10,000\vunit\ wide unidentified emission feature at $\sim$8600\,\AA, 
possibly suggestive of a terminal explosive event \citep{kasliwal11a,kasliwal12}. In 
general, though, the highest velocities in CK~Vul are considerably 
higher than those in LRNe.

The peak absolute luminosity of CK~Vul is also much higher than 
the Galactic LRNe, the brightest of which was V838~Mon with 
$M_{V} = -9.8$~mag \citep{bond03}. \cite{kasliwal12} has shown 
that there exists a group of objects with luminosities that are 
intermediate between the brightest novae and faintest supernovae
\citep[i.e. with peak $M_{V}$ between --9 and --14,][Fig.~1]{kasliwal12}. 
\cite{soker11} have termed these Intermediate-Luminosity Optical 
Transients (ILOTs). In the literature ILOTs  are interchangeably called Intermediate-Luminosity 
Red Transients or LRNe. V838~Mon, M31~RV, and M85OT are all proposed members 
of this group. It is suggested \citep{kasliwal12} that the less luminous of these 
outbursts result from mergers of main sequence stars: ``mergeburst'' 
events \citep{soker07}. 

The strongest evidence for a mergeburst is the case of V1309~Sco,
where a  decaying orbital period prior to eruption was clearly 
detected \citep{tylenda11}, indicating the components of the binary 
were spiraling into each other. However, V1309~Sco is estimated to have 
a peak $M_{V}$ of only $\sim-6$~mag (although the distance 
to the object is uncertain).

To explain the outbursts of the more luminous ILOTs, alternative  
routes are needed. One suggested channel is electron capture in an 
O-Ne-Mg core in extreme asymptotic giant branch (eAGB) stars 
\citep{kasliwal12}. The eAGB stars are massive AGB stars 
\citep[5-10\Msun,][]{soker11}. Alternatively, 
\citeauthor{soker11} discuss the possibility that such outbursts could 
be related to major instabilities of the envelopes of eAGB stars  
leading to ejections of their envelopes. The focus on eAGB stars 
arises because the mid-IR progenitors of two of the nearest
extragalactic ILOTs, NGC300-OT and SN 2008S, are at the extremely 
luminous and red ends of the AGB branch 
\citep{prieto08, thompson09, kasliwal12}. Other suggested mechanisms 
are luminous blue variable eruptions, 
asteroids falling onto  white dwarfs, accretion-induced collapses, 
peculiar core-collapse supernovae, and peculiar classical novae 
\citep[][and references therein]{kasliwal12}. 

We make no attempt to classify the nature of the CK~Vul
outburst (e.g., non-terminal Luminous Blue Variable, terminal events such
as SN~2008S and NGC 300 OT2008-1). Discriminating between these is 
challenging even for present day events, despite having diagnostic tools
such as panchromatic outburst spectra, archival images to identify the
nature of the progenitor 
\citep[e.g.,][]{smith10,adams16,jencson19,andrews20}, and in 
some cases the presence of a stellar remnant.

Any proposed  mechanism/model to account for the 
properties of the CK~Vul explosion must be able to account for  
its key features. These include the high peak outburst magnitude; 
the multiple peaks of the outburst; the 
sustained brightness of the outburst 
for a remarkably long period 
\citep[$\sim700$~days; see the reconstructed light curve in][]{shara85}; 
the high peak expansion velocities exceeding 2000\vunit; the formation 
and maintenance of jets, bubbles, and dust cocoons; and the 
production of the short-lived radio-nuclide \nucl{26}{Al} during its outburst
\citep{kaminski18a}.

\acknowledgments

\par
This Letter is based on observations obtained for program GN-2020B-Q-207
at the international Gemini Observatory, a program of NSF's NOIRLab, 
which is managed by the Association of Universities for Research in 
Astronomy (AURA) under a cooperative agreement with the National 
Science Foundation on behalf of the Gemini Observatory partnership: 
the National Science Foundation (United States), National Research 
Council (Canada), Agencia Nacional de Investigaci\'{o}n y Desarrollo 
(Chile), Ministerio de Ciencia, Tecnolog\'{i}a e Innovaci\'{o}n 
(Argentina), Minist\'{e}rio da Ci\^{e}ncia, Tecnologia, Inova\c{c}\~{o}es 
e Comunica\c{c}\~{o}es (Brazil), and Korea Astronomy and Space Science 
Institute (Republic of Korea).
DPKB is supported by a CSIR Emeritus Scientist grant-in-aid,
which is being hosted by the Physical Research Laboratory, Ahmedabad. 
We thank M. Hajduk for helpful comments on the manuscript.

\facilities{Gemini:Gillett}
\software{Figaro  \citep{figaro}, Gemini IRAF Package}





\end{document}